\documentstyle[12pt]{article}
\input epsf
\def \b{{\cal B}}
\def \beq{\begin{equation}}
\def \eeq{\end{equation}}
\def \beqn{\begin{eqnarray}}
\def \eeqn{\end{eqnarray}}
\def \cn{Collaboration}

\def \Kbar{\bar K}
\def \s{\sqrt{2}}

\def \st{\sqrt{3}}
\def \sx{\sqrt{6}}
\def \v#1#2{V_{#1#2}}

\begin{document}
\rightline{EFI-98-23}
\rightline{hep-ph/9806348}
\rightline{June 1998}
\bigskip
\bigskip
\centerline{{\bf RESCATTERING INFORMATION FROM $B \to K \bar K$ DECAYS}
\footnote{To be published in Phys.~Rev.~D.}}
\bigskip
\centerline{{\it Michael Gronau}
\footnote{Permanent address: Department of Physics, Technion -- Israel
Institute of Technology, Haifa 32000, Israel.}} 
\medskip
\centerline{and}
\medskip
\centerline{\it Jonathan L. Rosner}
\medskip
\centerline{\it Enrico Fermi Institute and Department of Physics}
\centerline{\it University of Chicago, Chicago, IL 60637}
\bigskip
\centerline{\bf ABSTRACT}
\medskip
\begin{quote}
Rescattering effects can modify the dependence on the weak phase $\gamma =
-{\rm Arg}(V^*_{ub}V_{ud}/V^*_{cb} V_{cd})$ of the ratio of rates for $B^{\pm}
\to K \pi^\pm$ and $B \to K^\pm \pi^\mp$.  A test for these effects based on
the processes $B^\pm \to K^\pm K$ has been suggested. It is pointed out that
the rates for the processes $B \to K^+ K^-$, which are expected to be {\it
dominated} by rescattering and for which considerably better experimental
bounds exist, are likely to provide a more stringent constraint on these
effects. 

\end{quote}
\medskip
\leftline{\qquad PACS codes:  12.15.Hh, 12.15.Ji, 13.25.Hw, 14.40.Nd}
% \newpage
\bigskip

\centerline{\bf I. INTRODUCTION}
\bigskip

The decays of $B$ mesons have provided useful insights into the pattern of weak
charge-changing transitions.  $B$ decays may serve as a new arena for the study
of CP violation, and may permit the direct measurement of phases of weak
couplings even when CP-violating effects have not been seen. Such is the case,
for example, when one compares rates for the decays $B^\pm \to K \pi^\pm$ and
$B \to K^\pm \pi^\mp$ \cite{RFa,GR,Wurt}.  (States without superscripts will
denote neutral mesons or their charge conjugates.)  In the simplest picture,
the decays $B^\pm \to K \pi^\pm$ are dominated by a ``penguin'' amplitude with
weak phase $\pi$, while the decays $B \to K^\pm \pi^\mp$ should contain a small
additional contribution from a ``tree'' amplitude with weak phase $\gamma$
\cite{RFa,GR,FM}.  The ratio \beq \label{eqn:Rdef} R \equiv \frac{\Gamma(B^0
\to K^+ \pi^-) + \Gamma(\bar B^0 \to K^- \pi^+)} {\Gamma(B^+ \to K^0 \pi^+) +
\Gamma(B^- \to \bar K^0 \pi^-)} \eeq was shown to provide useful information on
the relative importance of different weak subprocesses and hence on the weak
phase $\gamma = - {\rm Arg}(V^*_{ub} V_{ud}/V^*_{cb} V_{cd})$, especially when
complemented with information on CP-violating asymmetries such as parametrized
by the ratio 
\beq
A_0 \equiv \frac{\Gamma(B^0 \to K^+ \pi^-) - \Gamma(\bar B^0 \to 
K^- \pi^+)} {\Gamma(B^+ \to K^0 \pi^+) + \Gamma(B^- \to \bar K^0 \pi^-)}~~~.
\eeq

A number of recent works \cite{resc} have noted that rescattering effects, if
sufficiently large, could obviate the above results.  One test for such effects
\cite{Falk, RFb} relies on an SU(3) relation between their contributions in
$B^\pm \to K^\pm \pi$ and $B^\pm \to K^\pm K$ decays.  In the present paper we
analyze relations among such effects in {\it all} $B \to K \bar K$ charge
states. We find that the rates for the processes $B \to K^+ K^-$, which are
expected to be {\it dominated} by rescattering and for which better
experimental bounds exist, are likely to provide a more stringent constraint on
these effects.  We have previously emphasized the role of processes such as $B
\to K^+ K^-$ in evaluating the importance of rescattering \cite{BGR}. 

In Section II we recapitulate previous results on the determination of $\gamma$
through bounds \cite{FM} based on the ratio $R$ and through the combination of
$R$ with CP-violating asymmetry information as provided, for example, by $A_0$
\cite{GR}.  We discuss the criticisms raised in Refs.~\cite{resc} in Sec.~III,
where we also explain the relation between rescattering in $B^\pm \to K^\pm
\pi$ and $B^\pm \to K^\pm K$. Examples are given in Sec. IV of rescattering via
specific intermediate states, where relations among all charge states in $B \to
K \bar K$ occur.  We remark briefly about the effect of rescattering in
extracting the ratio of tree to penguin contributions in $B \to K^\pm \pi^\mp$
in Sec.~V, and summarize in Sec. VI. 

When studying rescattering effects we concentrate on two-body and
quasi-two-body intermediate states. It is likely that multiparticle
intermediate states play a dominant role in rescattering \cite{Donoghue}. We
will refer to such states only occasionally. Whereas quantitative studies of
rescattering effects via intermediate (quasi) two-body intermediate states are
crude and involve various dynamical assumptions \cite{Falk, Ger}, our present
qualitative discussion of such states will employ simple quark diagrams
demonstrating general conservation laws.
\bigskip

\centerline{\bf II.  REVIEW OF PREVIOUS RESULTS}
\bigskip

\leftline{\bf A.  Flavor-SU(3) decomposition}
\bigskip

The decays of $B$ mesons to two flavor-octet light pseudoscalar mesons are
characterized by 5 flavor-SU(3) invariant amplitudes \cite{ZepSWChau}. An
equivalent graphical description \cite{GHLR} in terms of an over-complete set
of six amplitudes displays the contributions in a manner which shows the flow
of flavor and color.  We use unprimed amplitudes to denote
strangeness-preserving $(\Delta S = 0)$ $b$ decays and primed amplitudes to
denote $b$ decays leading to one unit of net strangeness ($|\Delta S| = 1$). 

The amplitudes describing $B \to P_1 P_2$ decays, where $P_i$ denotes one of
the pseudoscalar SU(3)-octet mesons, are as follows: 

\begin{enumerate}

\item A {\it tree} amplitude $T$ ($T'$) involves the subprocess $\bar b \to
\bar u u \bar d$ ($\bar b \to \bar u u \bar s$) in which the $u \bar d$ ($u
\bar s$) produced by the weak current materializes into a single meson. 
Such a process is {\it color-favored} in the sense that it is of leading order
in an expansion of amplitudes in inverse powers of the number $N_c$ of quark
colors. 

\item A {\it color-suppressed} amplitude $C$ ($C'$) involves the same
subprocess as the corresponding tree amplitude, but the quark and antiquark
produced by the weak current end up in different mesons. This amplitude is
expected to be suppressed by a factor of $1/N_c$ with respect to the tree
amplitude. 

\item A {\it penguin} amplitude $P$ ($P'$) has the flavor structure $\bar b \to
\bar d$ ($\bar b \to \bar s$), where the light antiquark $\bar d$ ($\bar s$)
ends up in one of the final mesons, the spectator quark in the initial $B$ ends
up in the other, and a light quark-antiquark pair is produced in an
SU(3)-flavor-singlet state.  Electroweak penguins violate this last condition
and will be discussed separately. 

\item An {\it annihilation} amplitude $A$ ($A'$) involves the annihilation of
the $\bar b$ and the $u$ in a decaying $B^+$ into a weak current, which then
materializes into a pair of light pseudoscalar mesons. 

\item An {\it exchange} amplitude $E$ ($E'$) involves the subprocess $\bar b d
\to \bar u u$ ($\bar b s \to \bar u u$), where the initial light quark is in
the decaying particle, and thus contributes only to $B^0$ $(B_s)$ decays. 

\item A {\it penguin annihilation} amplitude $PA$ ($PA'$) involves the
annihilation of a $\bar b$ and $d$ ($\bar b$ and $s$) into a state with vacuum
quantum numbers, with subsequent production of a pair of light pseudoscalar
mesons. 

\end{enumerate}

These six amplitudes appear in 5 independent linear combinations, e.g., $C+T$,
$C-P$, $P+A$, $P+PA$, and $E+PA$, corresponding to the 5 SU(3) invariant
amplitudes. Since penguin processes involves loop diagrams with at least one
additional power of $\alpha_s$, they are expected to be modestly suppressed in
comparison with tree processes involving comparable sizes of
Cabibbo-Kobayashi-Maskawa (CKM) matrix elements. Since the last three processes
involve the participation of the spectator quark, they are expected to be
suppressed by a factor of $f_B/m_B$. The last process should be suppressed by
both effects. 

Electroweak penguin amplitudes \cite{EWP} involve no new flavor-SU(3)
structures, but require care in identifying weak phases.  They may be taken
into account by redefining each invariant amplitude to include an electroweak
penguin (EWP) contribution \cite{GHLREWP}, $t \equiv T + P^C_{\rm EW}$, $p
\equiv P - (1/3)P^C_{\rm EW}$, $c \equiv C + P_{\rm EW}$.  We shall ignore
these contributions \cite{GR, RFb} for the present discussion. 

Application of this SU(3) decomposition relies on associating certain weak
phases with some of the six amplitudes. $T~(T'),~C~(C'),~A~(A'),~E~(E')$ carry
the phase $\gamma$. Phases of penguin amplitudes are more involved and require
special care when rescattering corrections are considered. For instance, $P'$
is dominated by a weak phase $\pi$; however, rescattering corrections may
introduce a significant contribution with phase $\gamma$. While such
corrections do not affect the SU(3) decomposition, the interpretation of
invariant amplitudes can differ significantly from the naive one when
rescattering is important.  We shall give several concrete examples of this
circumstance. 

We shall discuss here only decays of nonstrange $B$ mesons into final states
consisting of $\pi \pi$, $K \pi$, and $K \bar K$.  SU(3)-breaking effects,
decays of $B_s$, and decays involving $\eta$ and $\eta'$ states have been
treated elsewhere \cite{GHLR,eta}. We quote in Tables I and II the
decomposition of the relevant decay amplitudes.  Overall signs are a
consequence of a specific phase convention for meson states \cite{GHLR}. 

\renewcommand{\thetable}{\Roman{table}}
\begin{table}
\caption{Decomposition of $\Delta S = 0$ $B \to PP$ amplitudes in terms of
SU(3) invariant amplitudes.}
\begin{center}
\begin{tabular}{r c c c c c c} \hline \hline
Decay~~~ & $T$ & $C$ & $P$ & $E$ & $A$ & $PA$ \\ \hline
$B^+ \to \pi^+ \pi^0$ & $-1/\s$ & $-1/\s$ & 0 & 0 & 0 & 0 \\
$K^+ \bar K^0$ & 0 & 0 & 1 & 0 & 1 & 0 \\
 & & & & & & \\
$B^0 \to \pi^+ \pi^-$ & $-1$ & 0 & $-1$ & $-1$ & 0 & $-1$ \\
$\pi^0 \pi^0$ & 0 & $-1/\s$ & $1/\s$ & $1/\s$ & 0 & $1/\s$ \\
$K^+ K^-$ & 0 & 0 & 0 & $-1$ & 0 & $-1$ \\
$K^0 \bar K^0$ & 0 & 0 & 1 & 0 & 0 & 1 \\ \hline \hline
\end{tabular}
\end{center}
\end{table}

\begin{table}
\caption{Decomposition of $B \to K \pi$ amplitudes in terms of
SU(3) invariant amplitudes.}
\begin{center}
\begin{tabular}{r c c c c c c} \hline \hline
Decay~~~ & $T'$ & $C'$ & $P'$ & $E'$ & $A'$ & $PA'$ \\ \hline
$B^+ \to K^0 \pi^+$ & 0 & 0 & 1 & 0 & 1 & 0 \\
$K^+ \pi^0$ & $-1/\s$ & $-1/\s$ & $-1/\s$ & 0 & $-1/\s$ & 0 \\
 & & & & & & \\
$B^0 \to K^+ \pi^-$ & $-1$ & 0 & $-1$ & 0 & 0 & 0 \\
$K^0 \pi^0$ & 0 & $-1/\s$ & $1/\s$ & 0 & 0 & 0 \\ \hline \hline
\end{tabular}
\end{center}
\end{table}

\bigskip

\leftline{\bf B.  Status of data}
\bigskip

The CLEO Collaboration \cite{CLEOPP} has presented evidence for several of the
decay modes listed in Tables I and II, and upper limits for others.  The
branching ratios are summarized in Table III.  We also quote our own estimates
\cite{DGRPRL} on the basis of the SU(3) decomposition in Tables I and II and an
estimate of the magnitude of invariant amplitudes. We note that these
estimates, based on measured $B \to K \pi$ and $B \to \pi \pi$ rates as input,
neglect SU(3) breaking effects and ignore interference between different terms.
These branching ratios will be useful when we come to discuss the contributions
of various hadronic states to rescattering processes.  We have ignored possible
CP-violating effects, assuming equal rates for processes and their
charge-conjugates. 
\bigskip

\begin{table}
\caption{Branching ratios ${\cal B}$ for $B \to PP$ decays, in units of
$10^{-5}$.  Experimental upper limits are 90\% c.l. including systematic
errors.  Theoretical predictions are based on $T$ ($T'$) and $P$ ($P'$)
contributions only, and interference between these two is ignored. Predictions
are the same for charge-conjugated states.} 
\begin{center}
\begin{tabular}{r c c} \hline \hline
Decay & ${\cal B}$ (Ex) & ${\cal B}$ (Th) \\ \hline
$B^+ \to \pi^+ \pi^0$ & $< 2.0$ & $0.4 \pm 0.2$ \\
$K^+ \bar K^0$ & $< 2.1$ & $0.08 \pm 0.02$ \\
$K^0 \pi^+$ & $2.3^{+1.1}_{-1.0} \pm 0.3 \pm 0.2$ & $1.6 \pm 0.4$ \\
$K^+ \pi^0$ & $< 1.6$ & $0.8 \pm 0.2$ \\
 & & \\
$B^0 \to \pi^+ \pi^-$ & $< 1.5$ & $0.9 \pm 0.4$ \\
$\pi^0 \pi^0$ & $< 0.93$ & $0.04 \pm 0.01$ \\
$K^0 \bar K^0$ & $< 1.7$ & $0.08 \pm 0.02$ \\
$K^+ K^-$ & $< 0.43$ & (a) \\
$K^+ \pi^-$ & $1.5^{+0.5}_{-0.4} \pm 0.1 \pm 0.1$ & $1.6 \pm 0.4$ \\
$K^0 \pi^0$ & $< 4.1$ & $0.8 \pm 0.2$ \\ \hline \hline
\end{tabular}
\end{center}
\leftline{(a) No $T$ or $P$ contributions}
\end{table}

\leftline{\bf C.  Fleischer-Mannel bound}
\bigskip

The predictions of Table III for the decays $B^+ \to K^0 \pi^+$ and $B^0 \to
K^+ \pi^-$ are based on the assumption that the $|P'|^2$ contribution is the
only source of $B^+ \to K^0 \pi^+$ and is dominant in $B^0 \to K^+ \pi^-$,
where a very small $T'$ contribution is also expected. The equality of the two
rates is certainly consistent with present data.  However, Fleischer and Mannel
\cite{FM} have pointed out that if the two rates differ significantly, with $R
<1$ [see Eq.~(\ref{eqn:Rdef})] as suggested by the central value $R = 0.65 \pm
0.40$, one can obtain a useful upper bound on $\sin \gamma$. 

If we ignore a small $A'$ contribution, the amplitude for $B^+ \to K^0 \pi^+$
may be written 
\beq \label{eqn:B+}
A(B^+ \to K^0 \pi^+) = - |P'|~~~,
\eeq
where we have taken account of the weak phase Arg$(V_{tb}^* V_{ts}) = \pi$, and
have assumed that the phase of the $\bar b \to \bar s$ penguin amplitude is
dominated by the top quark contribution.  Nothing changes in this discussion if
one adds contributions from an internal c-quark with weak phase Arg$(V_{cb}^*
V_{cs}) = 0$, as has been suggested recently \cite{Ciu}.  An immediate test of
the dominance of this process by a single weak phase is the equality of the
rates for $B^+ \to K^0 \pi^+$ and $B^- \to \bar K^0 \pi^-$ \cite{GR,resc}. 

The amplitudes for $B^0 \to K^+ \pi^-$ and $\bar B^0 \to K^- \pi^+$ are given,
under similar assumptions (one uses isospin symmetry to relate the penguin
amplitudes in neutral and charged $B$ decays to $K\pi$ states), by 
\beq \label{eqn:B0}
A(B^0 \to K^+ \pi^-) = |P'| - |T'|e^{i \delta} e^{i \gamma}~~,~~~
A(\bar B^0 \to K^- \pi^+) = |P'| - |T'|e^{i \delta} e^{-i \gamma}~~~,
\eeq
where $\delta$ is a final-state phase difference between penguin and tree
amplitudes.  The ratio $R$ defined in Eq.~(\ref{eqn:Rdef}) is then 
\beq \label{eqn:bigR}
R = 1 - 2 r \cos \gamma \cos \delta + r^2~~~,
\eeq
where $r \equiv |T'/P'|$.  For fixed $R<1$ and any $r,\delta$ the minimum of
$|\cos \gamma| = |(R - 1 - r^2)/(2r \cos \delta)|$ occurs when $\cos \delta =
1$ and $r = (1-R)^{1/2}$, leading to the bound 
\beq \label{eqn:FM}
\sin^2 \gamma \leq R~~~.
\eeq

\leftline{\bf D.  Determination of $\gamma$}
\bigskip

If one knows $r$ in Eq.~(\ref{eqn:bigR}) and measures the CP-violating
asymmetry in $B \to K^\pm \pi^\mp$ decays one can solve for $\gamma$
\cite{RFa,GR,Wurt}.  Defining the pseudo-asymmetry 
\beq 
A_0 \equiv \frac{\Gamma(B^0 \to K^+ \pi^-) - \Gamma(\bar B^0 \to K^- 
\pi^+)}{\Gamma(B^+ \to K^0 \pi^+) + \Gamma(B^- \to \bar K^0 \pi^-)}~~~,
\eeq
one has $A_0 =  2 r \sin \delta \sin \gamma$, so \beq \label{eqn:Req} R = 1 +
r^2 \pm \sqrt{4 r^2 \cos^2 \gamma - A_0^2 \cot^2 \gamma}~~~. 
\eeq
This can be formally solved to give
$$
4r \sin \gamma = \pm \{ [(1+r)^2-(R+A_0)][(R-A_0)-(1-r)^2] \}^{1/2}
$$
\beq \label{eqn:gamma}
\pm \{ [(1+r)^2-(R-A_0)][(R+A_0)-(1-r)^2] \}^{1/2}~~~. 
\eeq
Estimates of $r$ include $0.16 \pm 0.06$ \cite{GR} and $0.20 \pm 0.07$
\cite{Wurt}.  A measurement of $\gamma$ to an accuracy of $\pm 10^\circ$ will
require $r$ to be known to $\pm 10\%$.  This error seems achievable \cite{GR}. 

The simplicity of this method depends on the assumption that the decay $B^+ \to
K^0 \pi^+$ is dominated by the $P'$ amplitude which has a single weak phase.
Other contributions from rescattering with a different weak phase would show up
as a CP-violating asymmetry in $B^+ \to K^0 \pi^+$ vs. $B^- \to \bar K^0 \pi^-$
decay rates \cite{resc}. Fleischer \cite{RFb} argues that a modified version of
the bound (\ref{eqn:FM}) can still be written, while rescattering effects might
prevent a sufficiently accurate determination of $r$. In the next two sections
we shall relate the rescattering contributions in $B^+ \to K^0 \pi^+$ to their
contributions in $B \to K \bar K$ decays, where of particular interest is $B^0
\to K^+ K^-$ which is dominated by rescattering. The question of 
rescattering effects on $r$ will be discussed in Sec.~V. 
\bigskip

\centerline{\bf III.  RESCATTERING EFFECTS}
\bigskip

\leftline{\bf A.  Diagrammatic representation}
\bigskip

The prediction that $\Gamma(B^+ \to K^0 \pi^+) = \Gamma(B^- \to \bar K^0
\pi^-)$ relies on the dominance of a single weak phase (that of the $P'$
amplitude).  In the absence of rescattering (we ignore small electroweak
penguin effects) and if an annihilation contribution $A'$ is as small as
expected \cite{GHLR}, $A(B^+ \to K^0 \pi^+) = A(B^- \to \bar K^0 \pi^-)$.
Moreover, rescattering contributions with a {\it different} weak phase than
that of $P'$ are needed in order to violate this relation. Rescattering
amplitudes from intermediate charm-anticharm states carrying the same isospin
and the same phase (mod $\pi$) as $P'$ do not affect the discussion of Secs. II
C and II D. 

Typical rescattering contributions to $B^+ \to K^0 \pi^+$ from intermediate
states of two charmless pseudoscalar mesons are illustrated in Fig.~1. We
consider only processes involving the $T'$ production amplitude for these
intermediate states, with the CKM structure $V_{ub}^* V_{us}$.  The weak phase
of this combination is $\gamma$, so rescattering from intermediate states
produced via the $T'$ amplitude can contribute to a CP-violating asymmetry in
$B^\pm \to \pi^\pm K$ decays. We omit for now contributions of the
color-suppressed $C'$ amplitude, which has the same weak phase as $T'$. The
contributions of Figs. 1(a) and 1(b) should be added coherently with a relative
$+$ sign, corresponding to the S-wave nature of the decay.  The contribution of
Fig.~1(c) may be related to those of Figs.~1(a) and (b) in some models (such as
Regge pole exchange) but is independent in general. 

\begin{figure}
\centerline{\epsfysize = 7 in \epsffile {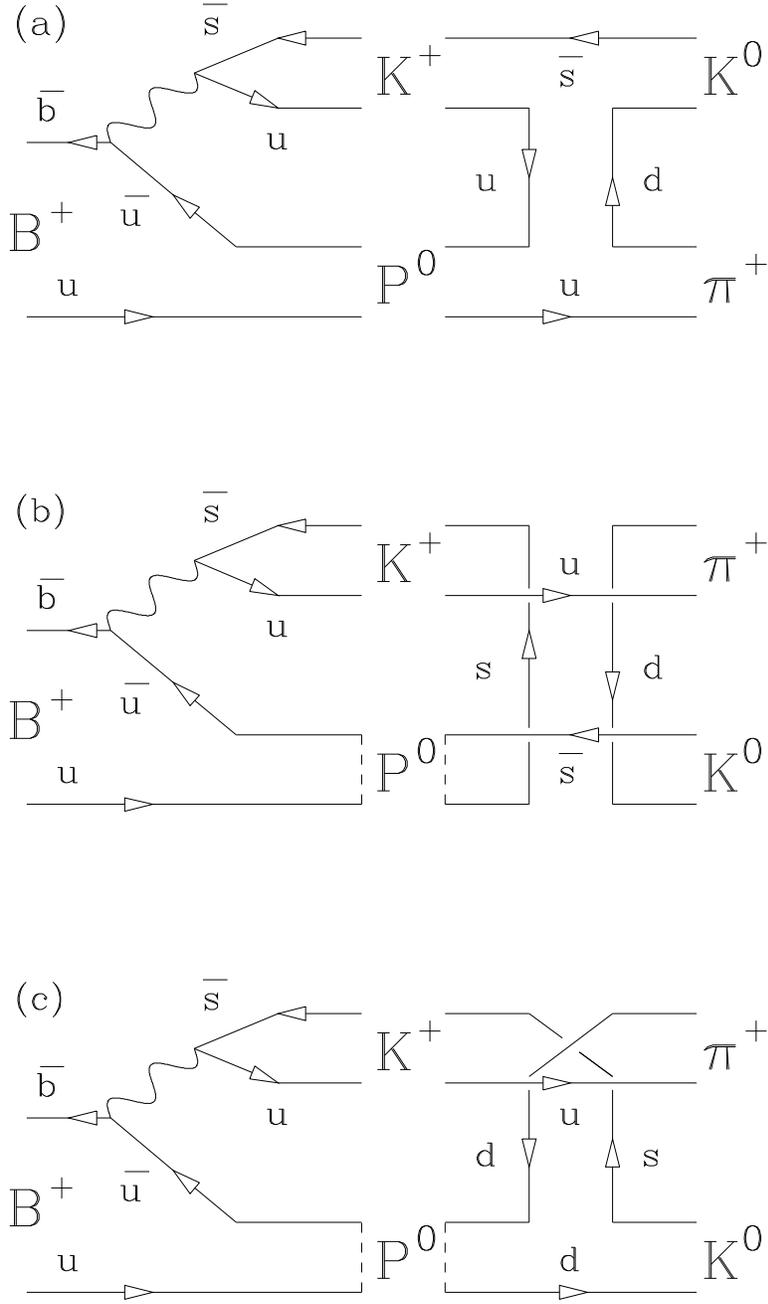}} \caption{Typical
rescattering contributions to $B^+ \to K^0 \pi^+$ from intermediate states of
two pseudoscalar mesons.  Here $P^0$ denotes $\pi^0, \eta, \eta'$.  (a)
Non-strange meson exchange with topology of $P'_u$ or $A'$, depending on how
quark lines in $P^0$ are connected; (b,c) strange meson exchange with topology
of $A'$. The dashed lines in (b,c) serve only to guide the eye in determining
the topology.} 
\end{figure}

The topology of quark lines in Fig.~1 illustrates the mixing of invariant
flavor-SU(3) amplitudes induced by rescattering.  Consider, for example,
Fig.~1(a).  Viewed as a diagram in which quark lines flow through meson
intermediate states from left to right, Fig.~1(a) has the topology of a $\bar b
\to \bar s$ penguin diagram in which a $u$ quark is the intermediate state in
the penguin amplitude.  We shall denote the corresponding amplitude by $P'_u$.
Similarly, $P'_{c,t}$ will denote penguin amplitudes for $\bar b \to \bar s$
with $c,t$ intermediate states.  A corresponding notation $P_{u,c,t}$ will
denote penguin amplitudes for $\bar b \to \bar d$ transitions. 

In the limit in which one sums over all meson intermediate states, one may
expect a form of quark-hadron duality in which Fig.~1(a) is just equivalent to
a short-distance $P'_{u}$ amplitude, expected to be smaller than $P'_{c,t}$ by
a factor $\vert V^*_{ub}V_{us}/V^*_{cb}V_{cs}\vert$.  This would involve a
cancellation of contributions reminiscent of that invoked \cite{Georgi} to
suppress $D^0 - \bar D^0$ mixing. When certain intermediate states are more
important than others this duality could well be violated, leading to large
rescattering contributions \cite{resc}. Thus, it makes sense to explore the
contributions of the lowest-mass intermediate states to gain at least a {\it
qualitative} understanding of relations among rescattering contributions to
various processes. 

There is another way to connect quark lines entering and leaving the neutral
meson $P^0$ in Fig.~1(a).  One could join the $u$ and $\bar u$ on the left with
one another and the $u$ and $\bar u$ on the right with one another, making a
pair of ``hairpins'' on the left and right of $P^0$. Such a diagram would have
the topology of an ``annihilation'' diagram, since it is equivalent to the
initial $\bar b$ and $u$ annihilating one another.  This ``hairpin'' diagram is
the only one possible in the diagram of Figs.~1(b) and 1(c). 

In the limit in which mass differences among $\pi^0$, $\eta$, and $\eta'$ can
be neglected, and in which these states are orthogonal combinations of $u \bar
u$, $d \bar d$, and $s \bar s$, the sum of their contributions to $q_i \bar q_i
\to q_j \bar q_j$, $i \ne j$, should vanish. This is just the familiar nonet
symmetry associated with the Okubo-Zweig-Iikuza (OZI) rule.  It probably holds
less well for pseudoscalar mesons (which can mix strongly with gluonic
intermediate states) than for the vast majority of other mesons.  Thus, the
graphs of Figs.~1(b) and 1(c) (and hence the topology associated with the $A'$
amplitude) should be important only when intermediate states involving
pseudoscalar mesons play a major role in rescattering contributions. If we were
to replace the intermediate state $K^+ P^0$ in Fig.~1 by a pair of vector
mesons $K^{*+} V^0$, the diagrams of Figs.~1(b) and 1(c) should be highly
suppressed, since nonet symmetry is very good for vector mesons.  Lipkin has
stressed the importance of this feature for $B$ decays in other contexts
\cite{HJLnonet}. 
\bigskip

\leftline{\bf B.  Relation between rescatterings in $B \to K\pi$ and $B 
\to K
\bar K$} 
\bigskip

Several authors \cite{Falk,RFb} have noted an SU(3) relation between
contributions to rescattering in $B^+ \to K^0 \pi^+$ and $B^+ \to K^+ \bar
K^0$. The corresponding $\Delta S = 1$ and $\Delta S = 0$ low energy effective
Hamiltonians, describing the subprocesses $\bar b \to \bar s \bar q q$ and
$\bar b \to \bar d \bar q q$ ($q=u, d, s, c$), involve each two terms
multiplied by CKM factors $V^*_{cb}V_{cs},~V^*_{ub}V_{us}$ and
$V^*_{cb}V_{cd},~V^*_{ub}V_{ud}$, respectively. The two pairs of $\Delta S = 1$
and $\Delta S = 0$ effective operators are related to each other by a U-spin
reflection $d \leftrightarrow s$. The dominant (direct) amplitudes in $B^+ \to
K^0 \pi^+$ and $B^+ \to K^+ \bar K^0$, which are proportional to
$V^*_{cb}V_{cs}$ and $V^*_{cb}V_{cd}$ respectively, obey the hierarchy 
\beq
A_c(B^+ \to K^+ \bar K^0) = -\lambda A_c(B^+ \to K^0 \pi^+)~~~,
\eeq
where $\lambda = V_{us}/V_{ud} = 0.22$. On the other hand, the amplitudes in
$B^+ \to K^0 \pi^+$ and $B^+ \to K^+ \bar K^0$, which receive contributions
from the subprocesses $\bar b \to \bar u u \bar s$ and $\bar b \to \bar u u
\bar d$ followed by rescattering, are proportional to $V^*_{ub}V_{us}$ and
$V^*_{ub}V_{ud}$, respectively, and obey the opposite hierarchy 
\beq \label{eqn:hier}
A_u(B^+ \to K^+ \bar K^0) = {1\over \lambda}A_u(B^+ \to K^0 \pi^+)~~~.
\eeq
This relation is expected to hold between the amplitudes $P_u + A$ and $P'_u +
A'$ in any description of rescattering which respects flavor SU(3). Examples
will be given in the next section. 

Thus, the ratio $A_u/A_c$ of amplitudes with different weak phases describing
rescattering and direct decays in $B^+ \to K^+ \bar K^0$ should be about
$-1/\lambda^2$ times larger than the corresponding ratio in $B^+ \to K^0
\pi^+$. This makes $B^+ \to K^+ \bar K^0$ particularly sensitive to
rescattering effects of this kind. We argued in Ref.~\cite{GR} that $A_u/A_c$
might be as large as unity in $B^+ \to K^+ \bar K^0$, raising the predicted
rate by as much as a factor of about 4.  This could lead to a prediction ${\cal
B}(B^+ \to K^+ \bar K^0) \simeq (2 \div 4) \times 10^{-6}$ instead of the value
$(8 \pm 2) \times 10^{-7}$ quoted in Table III. The corresponding ratio of
amplitudes with different weak phases in $B^+ \to K^0 \pi^+$ could then be as
large as $\lambda^2 \simeq 0.05$, sufficient to prevent a very useful
determination of $\gamma$. Fleischer \cite{RFb} has used larger rescattering
effects (via charmless intermediate states), and argued that conceivable values
of the squares of these amplitude ratios could be a factor of 5 above our
estimates, leading to possible values of ${\cal B}(B^+ \to K^+ \bar K^0)$ as
large as $2 \times 10^{-5}$.  This is not in conflict with any current
experimental bound (see Table III). However, in the next Section we shall show
that, at least in a few illustrative examples of intermediate rescattering
states, one expects similar or larger values for ${\cal B}(B^0 \to K^+ K^-)$,
for which a much better upper experimental limit ($< 4.3 \times 10^{-6}$)
exists. 

We will study only rescattering via charmless intermediate states, although
some rescattering could also be due to states involving charm-anticharm. Our
purpose is mainly to show that such final state interactions in $B^0 \to K^+
K^-$ are as important as in $B^+ \to K^+ \bar K^0$, which in turn are enhanced
by factor $1/\lambda$ relative to those in $B^+ \to K^0 \pi^+$ affecting the
determination of $\gamma$. Final state interaction via charm-anticharm
intermediate states obey the opposite hierarchy (10) and do not affect the
measurement of $\gamma$ as explained in Sec. II. 

\bigskip

\centerline{\bf IV.  RELATIONS AMONG RESCATTERING AMPLITUDES IN $B \to K
\bar K$} 
\bigskip

Before discussing specific intermediate states, let us comment briefly on 
possible contributions from charm-anticharm states, such as $D^+ D^-$.  
\bigskip

\leftline{\bf A.  $\pi \pi$ and  $\pi \eta$ intermediate states}
\bigskip

The dominant direct contributions to $B^0 \to K^0 \bar K^0$ and $B^+ \to K^+
\bar K^0$ are expected to arise from the penguin amplitude $P$ and to lead to a
branching ratio for each process of $(8 \pm 2) \times 10^{-7}$, as noted in
Table III.  The direct contributions to the decay $B^0 \to K^+ K^-$ are only an
exchange ($E$) and a penguin annihilation ($PA$) amplitude and thus are
expected to be considerably smaller. On the other hand, the (color-favored)
decays $B^0 \to \pi^+ \pi^-$ and $B^+ \to \pi^+ \pi^0$ are expected to have
branching ratios of about $8 \times 10^{-6}$ and $4 \times 10^{-6}$,
respectively.  One might expect rescattering from these states into $K \bar K$
to be of some importance. 

The decays $B \to \pi \pi$ can only populate two-pion states of isospin $I=0$
and $I=2$ by virtue of Bose statistics.  The final $K \bar K$ states can have
only $I=0$ and $I=1$.  Consequently, rescattering from $\pi \pi$ states must
lead uniquely to an $I=0$ final $K \bar K$ state, with the consequence 
\beq \label{eqn:pipiresc}
A(B^0 \to \pi \pi \to K^+ K^-) = - A(B^0 \to \pi \pi \to K^0 \bar K^0)~~,~~~
A(B^+ \to \pi \pi \to K^+ \bar K^0) = 0
\eeq
independent of any detailed mechanisms.  In particular, this relation holds in
the presence of each separate contribution to $B \to \pi \pi$, i.e., $C$ and
$P$ as well as the dominant $T$. 

To illustrate how graphical contributions satisfy the relations
(\ref{eqn:pipiresc}), consider Figs.~2 and 3 which illustrate the rescattering
into $K \bar K$ from the color-favored $T$ contribution to $B \to \pi \pi$. The
contributions of Figs.~2(a) and 2(b) are equal and opposite, with the negative
relative sign coming from the convention adopted for meson states. In terms of
invariant SU(3) amplitudes, however, Fig.~2(a) has the topology of a $P_u$
amplitude, while Fig.~2(b) has the topology of $E$.  The hierarchy of invariant
amplitudes noted in \cite{GHLR,GHLREWP} thus is strongly affected if
rescattering is important. 

If $P^0$ in Figs.~3 is taken to denote a $\pi^0$, the contributions from
Figs.~3(a) and 3(b) exactly cancel one another as a result of the opposite
relative signs of the $u \bar u$ and $d \bar d$ components of the $\pi^0$,
while Fig.~3(c) does not enter into the calculation at all. Note that whereas
Fig.~3(a) has the topology of a $P_u$ or $A$ amplitude (depending on how the
quark lines entering and leaving the $\pi^0$ are connected with one another),
Fig.~3(b) has the topology of $A$. 

The U-spin relation mentioned in Sec.~III B cannot be applied if one considers
only intermediate $\pi \pi$ contributions to $B \to K \bar K$, since $\pi^0 =
(d \bar d - u \bar u)/\s$ transforms under $d \leftrightarrow s$ into $(s \bar
s - u \bar u)/\s = (\st \eta_8 + \pi^0)/2$.  Here $\eta_8$ denotes the
flavor-octet state $\eta_8 \equiv (2 s \bar s - u \bar u - d \bar d)/\sx$. One
should thus consider both $K^+ \pi^0$ and $K^+ \eta_8$ intermediate-state
contributions to $B^+ \to K^0 \pi^+$ (Fig.~1), and hence, for self-consistency,
also $\pi^+ \eta_8$ contributions to $B^+ \to K^+ \bar K^0$ (Fig.~3). The
diagram of Fig.~3(c) must then be included for $B^+ \to K^+ \bar K^0$.  I t is
equivalent to that of Fig.~1(c) but with the substitution $d \leftrightarrow s$
everywhere. Ignoring the mass difference between the $\pi^0$ and $\eta_8$, one
confirms Eq.~(\ref{eqn:hier}): 
\beq
A(B^+ \to [K^+ \pi^0, K^+ \eta_8] \to K^0 \pi^+) = \lambda A(B^+ \to [\pi^+
\pi^0, \pi^+ \eta_8] \to K^+ \bar K^0)~~~. 
\eeq

\begin{figure}
\centerline{\epsfysize = 4.5 in \epsffile {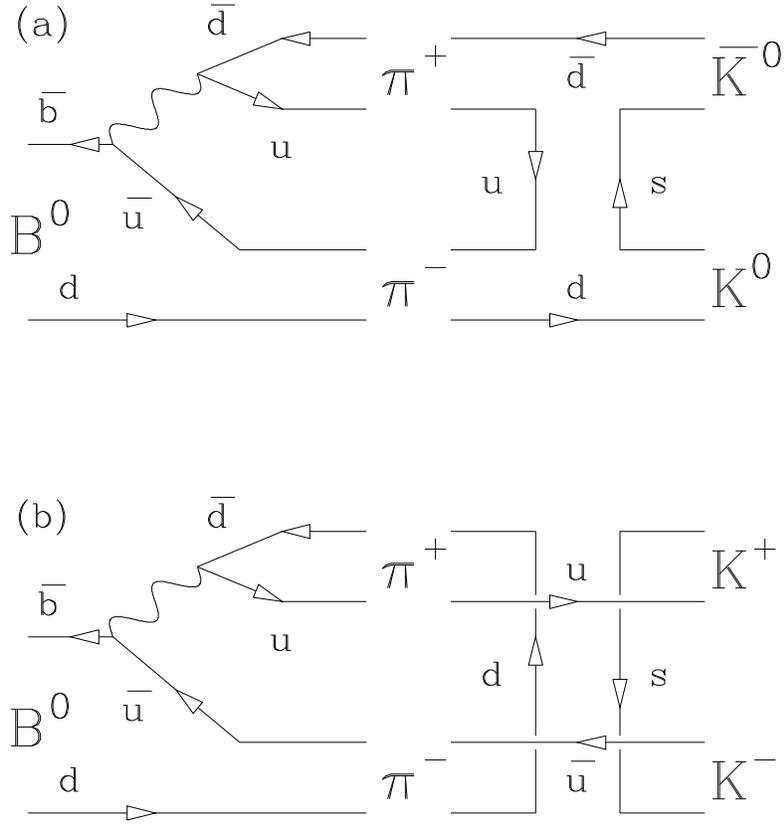}}
\caption{Rescattering contributions to $B^0 \to K \bar K$ from
$\pi^+ \pi^-$ intermediate states.  (a) $B^0 \to K^0 \bar K^0$ (topology
of $P_u$); (b) $B^0 \to K^+ K^-$ (topology of $E$).}
\end{figure}

\begin{figure}
\centerline{\epsfysize = 7 in \epsffile {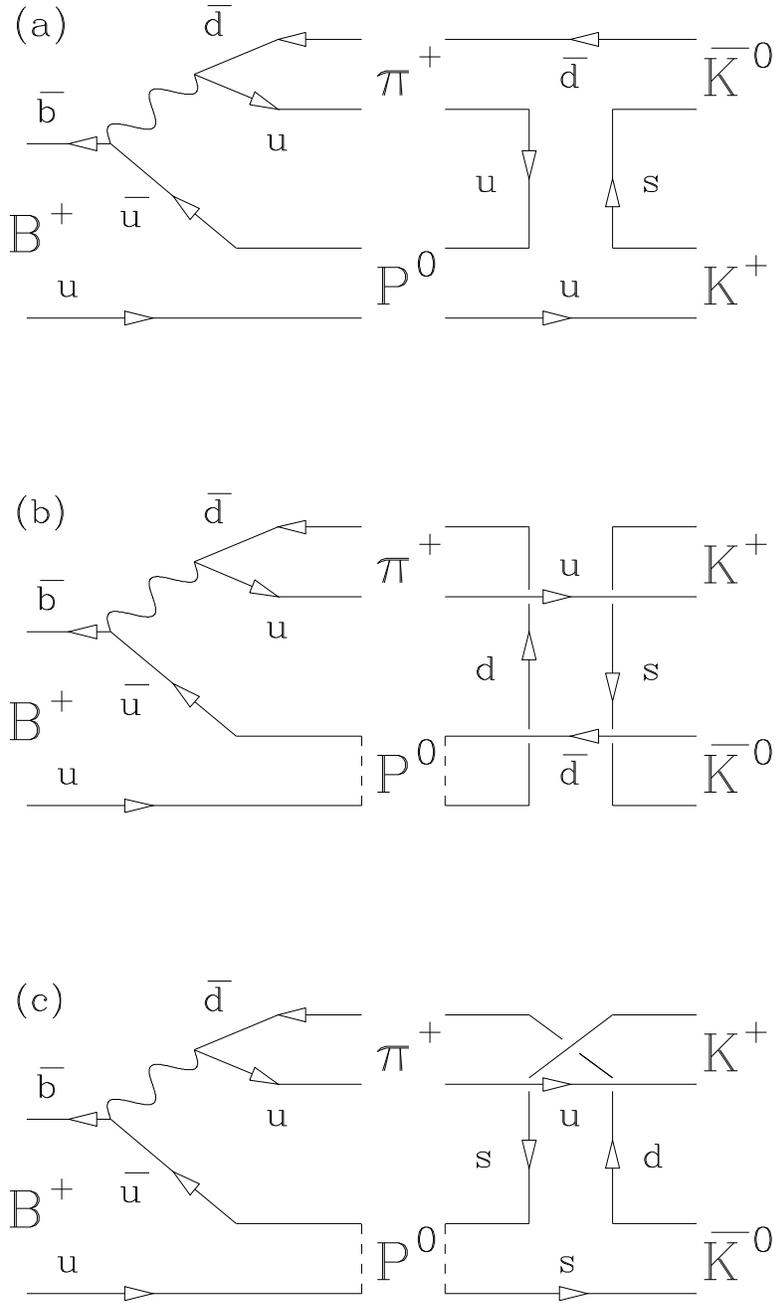}}
\caption{Rescattering contributions to $B^+ \to K^+ \bar K^0$ from $\pi^+ P^0$
intermediate states.  (a) Topology of $P_u$ or $A$; (b,c) topology of $A$.  The
contributions (a) and (b) must cancel one another exactly for $P^0 = \pi^0$
since $\pi^+ \pi^0$ in an S-wave has isospin $I=2$ while $K^+ \bar K^0$ in an
S-wave has $I=1$.} 
\end{figure}

Within a specific model of Regge pole exchange involving just exchange of the
leading strange vector and tensor meson trajectories \cite{Falk, Ger}, the
uncrossed graphs of Figs.~3(a,b) and the crossed graph of Fig.~3(c) are related
to one another by crossing symmetry \cite{dd}.  The graphs of Figs. 3(a) and
3(b) give equal amplitudes after S-wave projection.  [Note that the final
particles are interchanged in the two graphs, as in Figs.~1(a) and 1(b).] The
amplitude for an uncrossed graph in Fig.~3(a) has a phase $-e^{-i \pi
\alpha(t)}$, while the amplitude for an uncrossed graph in Fig.~3(b) has a
phase $-e^{-i \pi \alpha(u)}$, before S-wave projection.  Here $t \equiv
(p_{\bar K^0} - p_{\pi^+})^2$, $u \equiv (p_{K^+} - p_{\pi^+})^2$.  The
corresponding crossed graph in Fig.~3(c) has a phase $-1$ relative to the first
two before S-wave projection.  Here $\alpha$ denotes the exchange-degenerate
vector and tensor kaon trajectories, with $\alpha(0) \simeq 0.32$ \cite{Had87}.
One finds 
\beq
A(B^+ \to [\pi^+ \pi^0, \pi^+ \eta_8] \to K^+ \bar K^0) = - \frac{1}{3}
(1 + \Delta)A(B^0 \to \pi^+ \pi^- \to K^+ K^-)~~~,
\eeq
where $\Delta$ is the ratio of the S-wave projection of a crossed graph to the
S-wave projection of an uncrossed graph.  Unless $|\Delta|$ is much greater
than 1, we expect that the rescattering amplitude for $B^+ \to K^+ \bar K^0$,
assuming just $\pi^+ \pi^0$ and $\pi^+ \eta_8$ intermediate states, should be
smaller in magnitude than that of the neutral $B$ into $K^0 \bar K^0$ or $K^+
K^-$. 

We should remark parenthetically that the use of Regge pole models to estimate
S-wave scattering amplitudes for light mesons with c.m. energies of more than 5
GeV is highly dubious.  Regge pole exchanges are probably valid mainly for {\it
peripheral} partial waves, i.e., orbital angular momenta $l$ corresponding to
impact parameters $b \simeq l/k \simeq 1$ fm, where $k \simeq 2.6$ GeV$/c
\simeq 13$ fm$^{-1}$ is the c.m. 3-momentum.  Thus for c.m. energies
corresponding to those in $B$ decays to a pair of light mesons, peripheral
partial waves are of order $l \simeq 13$, whereas the central partial waves are
likely to be highly subject to absorption (or effects of Regge {\it cuts})
\cite{HH}. Consequently, we are not able to place too much stock in any
estimate of $\Delta$, in contrast to other considerations in the present paper
which are much less model-dependent. 

If one includes also $\pi \eta'$ intermediate states and neglects the mass
difference between the $\pi^0$, $\eta$, and $\eta'$, the diagrams of Figs.~3(b)
and 3(c) do not contribute.  One then finds 
\beq
A(B^+ \to [\pi^+ \pi^0, \pi^+ \eta, \pi^+ \eta'] \to K^+ \bar K^0) = 
- A(B^0 \to \pi^+ \pi^- \to K^+ K^-)~~~,
\eeq
and hence equal rescattering rates for all three $B \to K \bar K$ processes.
So, depending on whether we consider just $\pi \pi$, also $\pi \eta$, or all
three of $\pi \pi,~\pi \eta$, and $\pi \eta'$ intermediate states, we obtain a
rescattering rate for $B^+ \to K^+ \bar K^0$ which is either zero, smaller
than, or equal to the rates for the other two $B \to K \bar K$ processes. 
\bigskip

\leftline{\bf B.  Vector meson intermediate states}
\bigskip

An important class of intermediate states more massive than $PP$ which
contribute to $B \to PP$ decays are $VV$, where $V$ denotes a vector meson.
(Angular momentum and parity conservation forbid rescattering of $VP$ states
into $PP$).  Branching ratios at a level of a few times $10^{-5}$ were obtained
for $B^0 \to \rho^+ \rho^-, B^+ \to \rho^+ \rho^0$ and $B^+ \to \rho^+ \omega$
in several model-dependent calculations \cite{Model}. The importance (and
possibly even dominance) of the corresponding $K^* \rho$ intermediate states in
rescattering into $K \pi$ final states has been considered recently
\cite{JinDu}. 

Since $\rho^0 = (d \bar d - u \bar u)/\s$ and $\omega = (d \bar d + u \bar
u)/\s$ are nearly degenerate, it is sufficient to work in the rotated basis
$V_u = (\omega - \rho^0)/\s$ and $V_d = (\omega +\rho^0)/\s$.  The diagrams
describing rescattering contributions to $B \to K \bar K$ from intermediate
vector-meson states produced by the dominant tree ($T$) contributions are shown
in Figs.~4 and 5. 

As in Fig.~2, the $\rho^+ \rho^-$ intermediate state provides equal and
opposite contributions to $B^0 \to K^0 \bar K^0$ [Fig.~4(a)] and $B^0 \to K^+
K^-$ [Fig.~4(b)].  Here the isospin argument of Sec.~IV A again applies.
Although the $I=1$ state of $\rho^+ \rho^-$ can be produced in the decay, since
it can be formed by coupling the spins of $\rho^+ \rho^-$ to $S=1$, their
orbital angular momenta to $L=1$, and $\vec{S} + \vec{L} \equiv \vec{J}$ to
$J=0$, it is forbidden by parity to couple to $K \bar K$ in an S-wave. We then
find 
\beq \label{eqn:rrresc}
A(B^0 \to \rho^+ \rho^- \to K^+ K^-) = - A(B^0 \to \rho^+ \rho^- \to K^0 
\bar
K^0)~~~.
\eeq

\begin{figure}
\centerline{\epsfysize = 4.5 in \epsffile {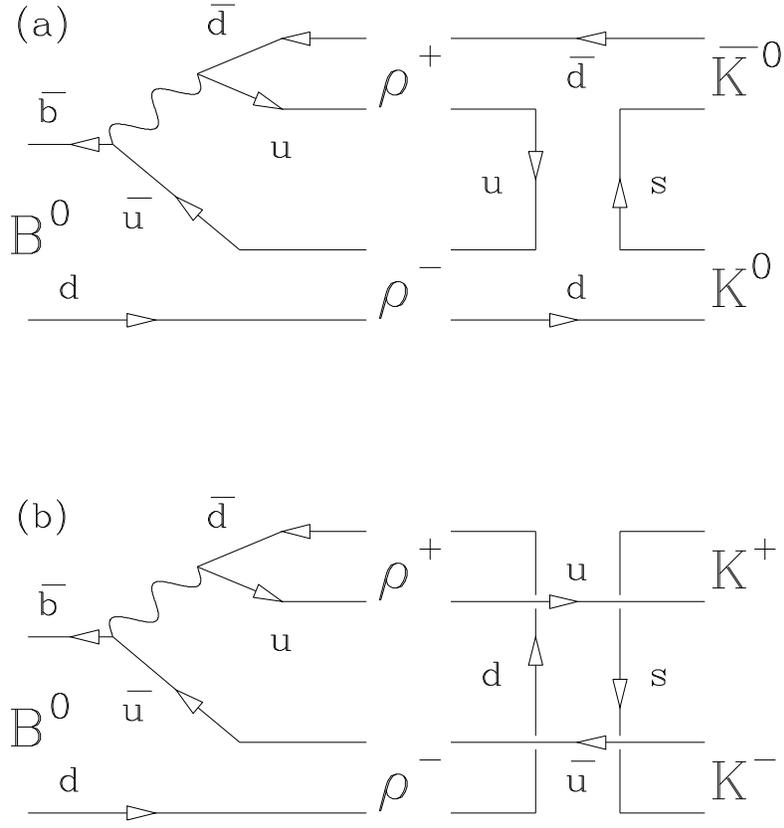}}
\caption{Rescattering contributions to $B^0 \to K \bar K$ from
$\rho^+ \rho^-$ intermediate states.  (a) $B^0 \to K^0 \bar K^0$ (topology
of $P_u$); (b) $B^0 \to K^+ K^-$ (topology of $E$).}
\end{figure}

\begin{figure}
\centerline{\epsfysize = 7 in \epsffile {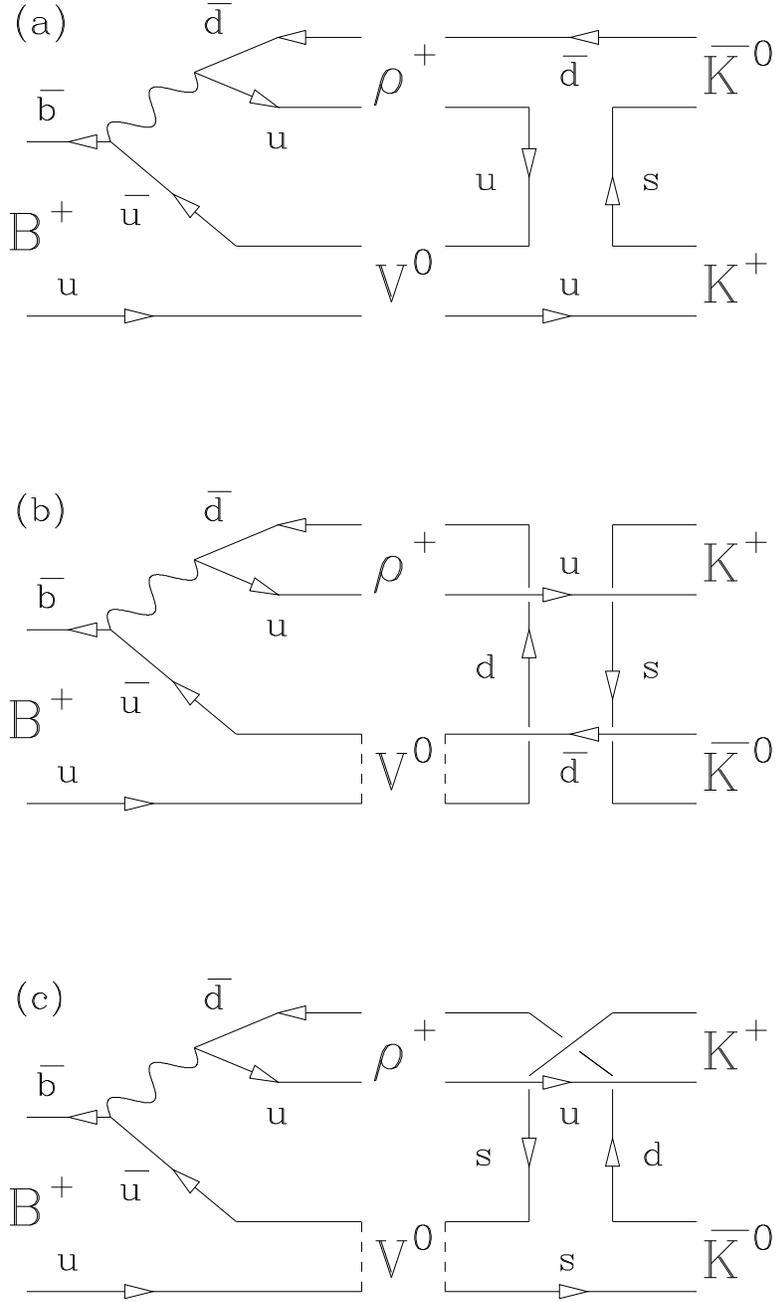}}
\caption{Rescattering contributions to $B^+ \to K^+ \bar K^0$ from $\rho^+ V^0$
intermediate states, where $V^0$ is a linear combination of $\rho^0$ and
$\omega$.  (a) Topology of $P_u$ or $A$; (b,c) Topology of $A$. Since $V^0$ is
produced as $V_u = u \bar u$ but must rescatter as $V_d = d \bar d$ (b) or $V_s
= s \bar s$ (c), the last two contributions must vanish.} 
\end{figure}

The graphs of Figs.~4(a) and 5(a) are identical, and the contributions of the
graphs of Fig.~5(b) and 5(c) must vanish if the vector mesons respect nonet
symmetry and the OZI rule. This implies a simple relation: 
\beq \label{eqn:rrrel}
A(B^+ \to \rho^+ V^0 \to K^+ \bar K^0) = A(B^0 \to \rho^+ \rho^- \to K^0 
\bar
K^0) = -A(B^0 \to \rho^+ \rho^- \to K^+ K^-) ~~~.
\eeq
Thus, the rescattering due to two vector mesons produced via the color-favored
$T$ amplitude gives equal contributions for all three $B \to K \bar K$
processes. 

The U-spin relation of Sec.~III B is evident if we perform the interchange $d
\leftrightarrow s$ on the graphs of Fig.~5. The result are the graphs of
Fig.~1, in which $K^+P^0$ are replaced by $K^{*+}V^0$. Fig.~5(a) then describes
the decay $B^+ \to K^0 \pi^+$ via an induced $P_u$ contribution, while
Figs.~5(b) and 5(c) continue to give vanishing contributions to this process. 

If one includes color-suppressed contributions to vector-meson pair production,
the simple relations (\ref{eqn:rrrel}) no longer seem to hold.  However, one
expects these contributions to be relatively small. 
\bigskip

\leftline{\bf C. $a_1 \pi$ and related intermediate states}
\bigskip

The branching ratio of $B^0 \to a^+_1 \pi^-$ was estimated \cite{BSW} to be
similar to that of $B^0 \to \rho^+ \pi^-$, a few times $10^{-5}$. The $a_1 \pi$
intermediate states, produced by dominant tree ($T$) contributions with weak
phase $\gamma$, can therefore lead to significant rescattering amplitudes into
$K \bar K$ states. 

In this case, a simple relation among the rescattering amplitudes into the
three $K \bar K$ states follows from G-parity conservation. Since the G-parity
of $a_1 \pi$ is $+1$, and that of $K \bar K$ in a state of angular momentum $L$
and isospin $I$ is $(-1)^{L+I}$, an S-wave $K \bar K$ state into which $a_1
\pi$ states rescatter must be pure $I=0$. Therefore, 
\beq \label{eqn:a1piresc}
A(B^0 \to a_1 \pi \to K^+ K^-) = - A(B^0 \to a_1 \pi \to K^0 \bar 
K^0)~~,~~~
A(B^+ \to a_1 \pi \to K^+ \bar K^0) = 0
\eeq
Again, as in the case of intermediate $\pi \pi$ states, this relation can be
demonstrated using figures analogous to Figs. 2 and 3. 

The $I=0$ partners of the $\pi$ are $\eta$ and $\eta'$; those of the $a_1$ are
$f_1(1285)$ and $f_1(1420)$ or $f_1(1510)$ \cite{PDG}.  These states have even
G-parity and probably contribute in color-allowed rescattering processes
leading to $B^+ \to K^+ \bar K^0$.  As in the case of rescattering from $PP$ or
$VV$ intermediate states, the $K^+ \bar K^0$ mode is not likely to be greatly
suppressed in a practical calculation. Our purpose was was rather to show that
the $K^+ K^-$ mode is not likely to be {\it smaller} than the others when
rescattering from a small number of specific intermediate states is dominant. 
\bigskip

\leftline{\bf D.  Inclusive intermediate states}
\bigskip

We would like to draw a more general conclusion from the previous examples. The
generic case of neutral mesons in intermediate states is probably more
analogous to the case of Sec.~IV B, in which nonet symmetry is valid and
transitions $q_i \bar q_i \to q_j \bar q_j~(i \ne j)$ are forbidden. Then
Figs.~1(a), 3(a), and 5(a) are interpreted purely as $P_u$, and contributions
of Figs. 1(b,c), 3(b,c), and 5(b,c) should vanish.  Hence, one finds no
color-favored rescattering contributions to annihilation-type amplitudes.
(There will still be color-suppressed contributions from rescattering to these
processes.) Color-favored rescattering processes $B \to M_1 M_2 \to K \bar K$
($M_1$ and $M_2$ are light-quark mesons) involving the CKM factor $V_{ub}^*
V_{ud}$ will then contribute equal amplitudes in all three $B \to K \bar K$
decays, which we would describe as effective $P_u$ and $E$ contributions. 

As one sums over more and more intermediate states contributing to the
rescattering process and neglects meson mass differences, we would expect the
relations among different processes to be more and more accurately described by
amplitudes corresponding to quark graphs \cite{GHLR}.  This corresponds to a
notion of quark-hadron duality akin to that in $e+ e^- \to {\rm hadrons}$ or
$\tau \to \nu_\tau + {\rm hadrons}$. When the intermediate hadronic states are
broad and overlapping, an effective description in terms of quarks and gluons
should become a good approximation.  One then needs, of course, to incorporate
the free quarks into pairs of light pseudoscalar mesons, which requires the
introduction of form factors.  The invariant amplitudes introduced in
\cite{GHLR} and similar approaches take such form factors into account in a
flavor-SU(3)-invariant manner.  Rescattering contributions then are described
in terms of quarks and gluons as well, as illustrated by the examples in
Fig.~6. The final quarks, as before, have to be incorporated into hadrons. 

\begin{figure}
\centerline{\epsfysize = 4.5 in \epsffile {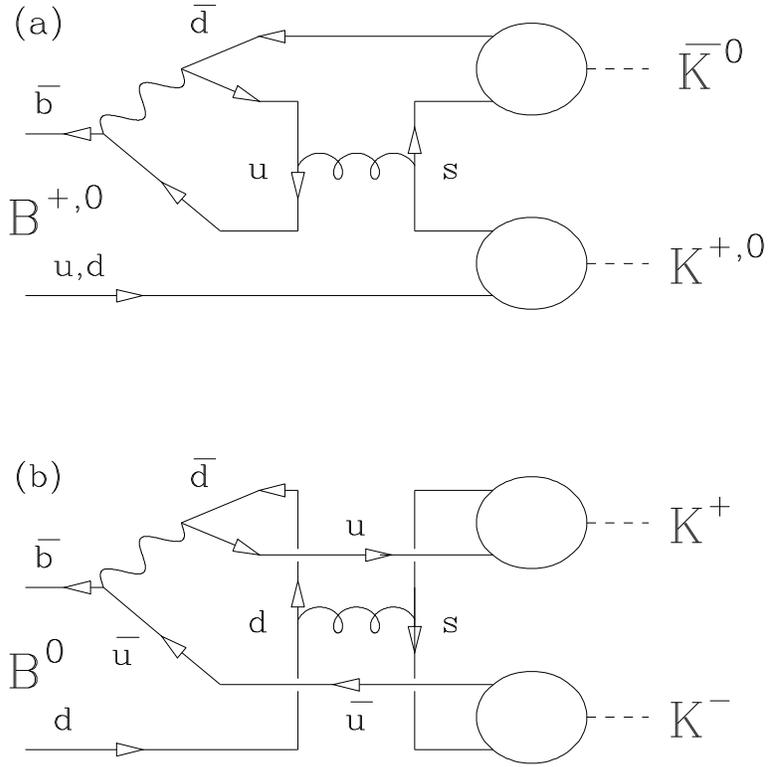}}
\caption{Examples of graphs contributing to the short-distance description of
rescattering in $B \to K \bar K$ processes.  The ovals denote form factors. 
(a) Processes with topology of a $P_u$ contribution; (b) process with topology
of an $E$ contribution.} 
\end{figure}

Contributions of $P_u$ and $P'_u$ graphs should be evaluable from a
short-distance point of view and are expected to be given roughly by
\cite{BFPu} $|P_u| \simeq |V^*_{ub}V_{ud}/V^*_{cb}V_{cd}||P|,~ |P'_u| \simeq
|V^*_{ub}V_{us}/V^*_{cb}V_{cs}||P'|$.  Here one has incorporated unknown form
factor information into the amplitude $|P'|$ which we have claimed is the
dominant contribution to observed $B \to K \pi$ decays. 

In the absence of significant long-distance effects the contributions of
$A~(A')$ and $E~(E')$-type graphs should contain a factor of $f_B/m_B$. It is
not clear how the form factors \cite{LB} in such graphs as Fig.~6(b) compare
with those in Fig.~6(a), however.  An explicit calculation is needed
\cite{Xing}; we expect it to be a more reliable guide to the magnitude of such
rescattering contributions than the popular Regge-pole analyses. 

As the hierarchy of amplitudes in terms of a graphical description becomes more
and more valid, one should then expect the prediction for the rate for $B^0 \to
K^+ K^-$ to drop significantly below that for $B^0 \to K^0 \bar K^0$ or $B^+
\to K^+ \bar K^0$. A rate for $B^0 \to K^+ K^-$ close to its present upper
experimental limit would indicate not only that rescattering contributions are
appreciable but that they violate the expected hierarchy of amplitudes. As we
have indicated in the two previous subsections, the decay rate for $B^0 \to K^+
K^-$ should be comparable to that for the other two $B \to K \bar K$ processes
if rescattering is an important contributor to the rates for these processes
and is dominated by a few specific intermediate states. 
\bigskip

\centerline{\bf V.  RESCATTERING AND TREE-PENGUIN AMPLITUDE RATIO}
\bigskip

In the first paper of Ref.~\cite{RFb} it was noted that rescattering could
affect the determination of the ratio $r = |T'/P'|$ which was needed to extract
the weak phase $\gamma$ from the ratio of $B^\pm \to K \pi^\pm$ and $B \to
K^\pm \pi^\mp$ rates.  This is true to some extent for the determination $r =
0.16 \pm 0.06$ \cite{GR}, which relied upon information from the decays $B \to
\pi^+ \pi^-$ and $B^\pm \to \pi^\pm \pi^0$.  In that determination it was
assumed that these processes were dominated by the color-favored amplitude $T$,
and that factorization could be used to relate $T$ to the corresponding
strangeness-changing amplitude $T'$. 

As noted in Ref.~\cite{GR}, a cleaner way to determine the $T$ amplitude in the
long run will be to use the semileptonic process $(B^0 \to \pi^- \ell^+
\nu_\ell)$, currently measured to have branching ratio \cite{CLEOsl} 
\beq \label{eqn:sl}
\b(B^0 \to \pi^- \ell^+ \nu_\ell) = (1.8 \pm 0.4 \pm 0.3 \pm 0.2) \times
10^{-4}~~~.
\eeq
When the spectrum for this process is well enough measured, one will use the
relation
\beq \label{eqn:dsl}
\Gamma(B^0 \to K^+ \pi^-)|_{\rm tree} = 6 \pi^2 f_K^2 |V_{us}|^2 a_1^2
\frac{d \Gamma(B^0 \to \pi^- \ell^+ \nu_\ell)}{dq^2} |_{q^2 = m_K^2}~~~
\eeq
to evaluate $T'$.

The key element in assuming that this factorization approach yields $T'$ arises
in the assumption that rescattering effects do not by themselves contribute a
significant $T'$ piece in $B \to K \pi$ decays. Note that $T'$ is defined as an
amplitude with weak phase $\gamma$. A typical rescattering contribution to $B^0
\to K^+ \pi^-$ carrying this phase is shown in Fig.~7(a). A corresponding
contribution to $B^+ \to K^0 \pi^+$ is shown in Fig.~7(b). An additional
contribution to $B^0 \to K^+ \pi^-$ of course comes from the elastic
intermediate state, whereas no such contribution with phase $\gamma$ occurs in
$B^+ \to K^0 \pi^+$. 

Using arguments as in Sec. IV, it can be seen that {\it inelastic} rescattering
is likely to be of comparable importance in $B^0 \to K^+ \pi^-$ and $B^+ \to
K^0 \pi^+$. For any inelastic channel leading to $K^+ \pi^-$ final state by a
diagram of type 7(a) there will be an isospin-related diagram of type 7(b), in
which a corresponding intermediate state rescatters to $B^+ \to K^0 \pi^+$.
Using this picture, the only difference between rescattering in the two
processes comes from the less important {\it elastic} channel which only
contributes to $B^0 \to K^+ \pi^-$. Similar elastic rescattering contributions
should affect $B^0 \to \pi^+ \pi^-$ or $B^+ \to \pi^+ \pi^0$. Their presence
would be manifested in a failure of factorization in the comparison of $B \to
\pi l \nu_l$ and color-favored $B \to \pi \pi$ decays. There are two ways to
gauge the importance of the major (inelastic) rescattering in $B^+ \to K^0
\pi^+$. One way is to look for rate enhancements in $B \to K \bar K$ as
discussed in Sec. IV. The other method \cite{GR, RFb} is by looking for a
CP-violating rate difference between $B^+ \to K^0 \pi^+$ and its charge
conjugate.   Thus, it appears that one will have satisfactory cross-checks of
the methods used to extract $r$ from $B$ decays. The method becomes
particularly simple if $B \to K \Kbar$ rates show no enhancement relative to
naive expectations, if no asymmetry is measured between $B^+ \to K^0 \pi^+$ and
its charge-conjugate, and if comparison of $B \to \pi l \nu_l$ with
color-favored $B \to \pi \pi$ decays supports factorization. 

\begin{figure}
\centerline{\epsfysize = 4.5 in \epsffile {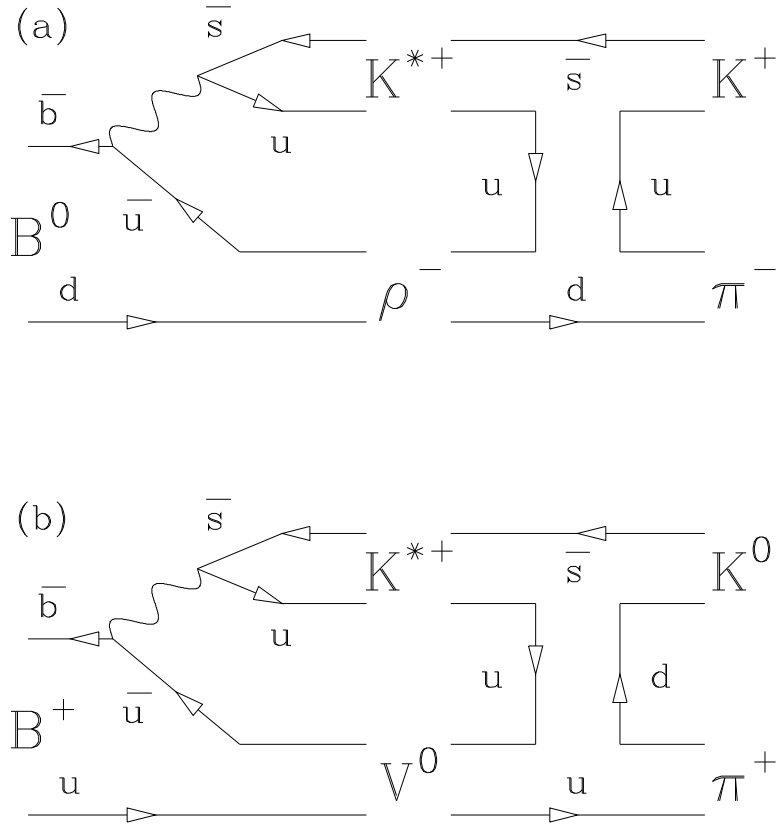}}
\caption{Rescattering contributions to (a) $B^0 \to K^+ \pi^-$ from $K^{*+}
\rho^-$ intermediate states, and (b) $B^+ \to K^0 \pi^+$ from $K^{*+} V^0$
intermediate states.} 
\end{figure}

\bigskip

\centerline{\bf VI.  SUMMARY}
\bigskip

We have discussed possible ambiguities in the determination of the weak phase
$\gamma$ through a comparison of $B^\pm \to K \pi^\pm$ and $B \to K^\pm
\pi^\mp$ decays. We have shown that satisfactory means exist for measuring the
effects of rescattering on these processes by studying the effects in $B\to K
\bar K$ decays. Rescattering effects in these processes are enhanced by
$1/\lambda^2$ relative to those in $B \to K \pi$. In particular, the decay $B^0
\to K^+ K^-$ is of great interest since it is dominated by rescattering
effects. We demonstrated a few cases in which the rescattering amplitude in
this process is expected to be as pronounced as in $B^+ \to K^+ \bar K^0$ and
$B^0 \to K^0 \bar K^0$. In the illustrative cases of $\pi \pi$ and $a_1 \pi$
intermediate states, rescattering into $K^+ K^-$ is allowed while rescattering
into $K^+ \bar K^0$ is forbidden by isospin and G-parity, respectively. Upper
limits on the rates of $B \to K^+ K^-$ can be used to set bounds on
rescattering effects in $B^\pm \to K \pi^\pm$, 
\beq
|P'_u/P'| \simeq \lambda \sqrt{{\Gamma(B^0\to K^+ K^-)+\Gamma(\bar B^0 \to 
K^+ K^-) \over \Gamma(B^+ \to K^0 \pi^+) + \Gamma(B^- \to \bar K^0 
\pi^-)}}~~~. 
\eeq

Whereas estimates of rescattering effects are rather crude and depend on
rescattering models (such as Regge-exchange \cite{Falk,Ger}), our present
considerations were model-independent once one assumed a dominant set of
intermediate states contributing to the rescattering. Our results were shown to
depend somewhat on the intermediate states through which rescattering occurs. 

In the absence of rescattering contributions, or when rescattering
contributions respect a hierarchy of amplitudes which predicts a suppression of
processes involving the spectator quark, the decays $B \to K^+ K^-$ are
expected to be highly suppressed. A very useful upper limit on the average
branching ratio of these processes would be $4\times 10^{-8}$, two orders of
magnitude below the present limit, which seems achievable in future experiments
\cite{Wein}. In this case the method we have proposed previously should be
sufficient for measuring $\gamma$ to a level of $10^\circ$ \cite{GR}. A more
modest limit, $4\times 10^{-7}$, would leave an uncertainty in $\gamma$ of the
order of a few tens of degrees. Conversely, an observation of these decay modes
may provide an early warning of the importance of rescattering effects, since
present experimental bounds on them are considerably more stringent than on
other modes expected to be enhanced by rescattering effects. 
\bigskip

\centerline{\bf ACKNOWLEDGEMENTS}
\bigskip

We thank A. Falk and F. W\"urthwein for discussions. This work was supported in
part by the United States -- Israel Binational Science Foundation under
Research Grant Agreement 94-00253/2 and by the United States Department of
Energy under Contract No. DE FG02 90ER40560. 
% \bigskip
\newpage

% Journal and other miscellaneous abbreviations for references
% Phys. Rev. D style
\def \ajp#1#2#3{Am.~J.~Phys.~{\bf#1}, #2 (#3)}
\def \apny#1#2#3{Ann.~Phys.~(N.Y.) {\bf#1}, #2 (#3)}
\def \app#1#2#3{Acta Phys.~Polonica {\bf#1}, #2 (#3)}
\def \arnps#1#2#3{Ann.~Rev.~Nucl.~Part.~Sci.~{\bf#1}, #2 (#3)}
\def \cmp#1#2#3{Commun.~Math.~Phys.~{\bf#1}, #2 (#3)}
\def \cmts#1#2#3{Comments on Nucl.~Part.~Phys.~{\bf#1}, #2 (#3)}
\def \corn93{{\it Lepton and Photon Interactions:  XVI International
Symposium, Ithaca, NY August 1993}, AIP Conference Proceedings No.~302,
ed.~by P. Drell and D. Rubin (AIP, New York, 1994)}
\def \cp89{{\it CP Violation,} edited by C. Jarlskog (World Scientific,
Singapore, 1989)}
\def \dpff{{\it The Fermilab Meeting -- DPF 92} (7th Meeting of the
American Physical Society Division of Particles and Fields), 10--14
November 1992, ed. by C. H. Albright \ite~(World Scientific, Singapore,
1993)}
\def \dpf94{DPF 94 Meeting, Albuquerque, NM, Aug.~2--6, 1994}
\def \efi{Enrico Fermi Institute Report No. EFI}
\def \el#1#2#3{Europhys.~Lett.~{\bf#1}, #2 (#3)}
\def \f79{{\it Proceedings of the 1979 International Symposium on Lepton
and Photon Interactions at High Energies,} Fermilab, August 23-29, 1979,
ed.~by T. B. W. Kirk and H. D. I. Abarbanel (Fermi National Accelerator
Laboratory, Batavia, IL, 1979}
\def \hb87{{\it Proceeding of the 1987 International Symposium on Lepton
and Photon Interactions at High Energies,} Hamburg, 1987, ed.~by W. Bartel
and R. R\"uckl (Nucl. Phys. B, Proc. Suppl., vol. 3) (North-Holland,
Amsterdam, 1988)}
\def \ib{{\it ibid.}~}
\def \ibj#1#2#3{~{\bf#1}, #2 (#3)}
\def \ichep72{{\it Proceedings of the XVI International Conference on High
Energy Physics}, Chicago and Batavia, Illinois, Sept. 6--13, 1972,
edited by J. D. Jackson, A. Roberts, and R. Donaldson (Fermilab, Batavia,
IL, 1972)}
\def \ijmpa#1#2#3{Int.~J.~Mod.~Phys.~A {\bf#1}, #2 (#3)}
\def \ite{{\it et al.}}
\def \jmp#1#2#3{J.~Math.~Phys.~{\bf#1}, #2 (#3)}
\def \jpg#1#2#3{J.~Phys.~G {\bf#1}, #2 (#3)}
\def \lkl87{{\it Selected Topics in Electroweak Interactions} (Proceedings
of the Second Lake Louise Institute on New Frontiers in Particle Physics,
15--21 February, 1987), edited by J. M. Cameron \ite~(World Scientific,
Singapore, 1987)}
\def \ky85{{\it Proceedings of the International Symposium on Lepton and
Photon Interactions at High Energy,} Kyoto, Aug.~19-24, 1985, edited by M.
Konuma and K. Takahashi (Kyoto Univ., Kyoto, 1985)}
\def \mpla#1#2#3{Mod.~Phys.~Lett.~A {\bf#1}, #2 (#3)}
\def \nc#1#2#3{Nuovo Cim.~{\bf#1}, #2 (#3)}
\def \nima#1#2#3{Nucl.~Instr.~Meth.~A {\bf#1}, #2 (#3)}
\def \np#1#2#3{Nucl.~Phys.~{\bf#1}, #2 (#3)}
\def \pisma#1#2#3#4{Pis'ma Zh.~Eksp.~Teor.~Fiz.~{\bf#1}, #2 (#3) [JETP
Lett. {\bf#1}, #4 (#3)]}
\def \pl#1#2#3{Phys.~Lett.~{\bf#1}, #2 (#3)}
\def \plb#1#2#3{Phys.~Lett.~B {\bf#1}, #2 (#3)}
\def \pr#1#2#3{Phys.~Rev.~{\bf#1}, #2 (#3)}
\def \pra#1#2#3{Phys.~Rev.~A {\bf#1}, #2 (#3)}
\def \prd#1#2#3{Phys.~Rev.~D {\bf#1}, #2 (#3)}
\def \prl#1#2#3{Phys.~Rev.~Lett.~{\bf#1}, #2 (#3)}
\def \prp#1#2#3{Phys.~Rep.~{\bf#1}, #2 (#3)}
\def \ptp#1#2#3{Prog.~Theor.~Phys.~{\bf#1}, #2 (#3)}
\def \rmp#1#2#3{Rev.~Mod.~Phys.~{\bf#1}, #2 (#3)}
\def \rp#1{~~~~~\ldots\ldots{\rm rp~}{#1}~~~~~}
\def \si90{25th International Conference on High Energy Physics, Singapore,
Aug. 2-8, 1990}
\def \slc87{{\it Proceedings of the Salt Lake City Meeting} (Division of
Particles and Fields, American Physical Society, Salt Lake City, Utah,
1987), ed.~by C. DeTar and J. S. Ball (World Scientific, Singapore, 1987)}
\def \slac89{{\it Proceedings of the XIVth International Symposium on
Lepton and Photon Interactions,} Stanford, California, 1989, edited by M.
Riordan (World Scientific, Singapore, 1990)}
\def \smass82{{\it Proceedings of the 1982 DPF Summer Study on Elementary
Particle Physics and Future Facilities}, Snowmass, Colorado, edited by R.
Donaldson, R. Gustafson, and F. Paige (World Scientific, Singapore, 1982)}
\def \smass90{{\it Research Directions for the Decade} (Proceedings of the
1990 Summer Study on High Energy Physics, June 25 -- July 13, Snowmass,
Colorado), edited by E. L. Berger (World Scientific, Singapore, 1992)}
\def \stone{{\it B Decays}, edited by S. Stone (World Scientific,
Singapore, 1994)}
\def \tasi90{{\it Testing the Standard Model} (Proceedings of the 1990
Theoretical Advanced Study Institute in Elementary Particle Physics,
Boulder, Colorado, 3--27 June, 1990), edited by M. Cveti\v{c} and P.
Langacker (World Scientific, Singapore, 1991)}
\def \yaf#1#2#3#4{Yad.~Fiz.~{\bf#1}, #2 (#3) [Sov.~J.~Nucl.~Phys.~{\bf #1},
#4 (#3)]}
\def \zhetf#1#2#3#4#5#6{Zh.~Eksp.~Teor.~Fiz.~{\bf #1}, #2 (#3) [Sov.~Phys.
- JETP {\bf #4}, #5 (#6)]}
\def \zpc#1#2#3{Zeit.~Phys.~C {\bf#1}, #2 (#3)}

\end{document}